\documentclass[a4paper, conference,10pt] {IEEEtran}
\usepackage[ left=0.64in, right=0.64in, top=0.75in, bottom=0.75in]{geometry}
\usepackage{amsmath}
\usepackage{amssymb}
\usepackage{bm}
\usepackage{color}
\usepackage{graphicx}

\usepackage[linesnumbered,ruled,lined]{algorithm2e}

\usepackage[outdir=./]{epstopdf}
\usepackage{caption}
\usepackage{lipsum}
\usepackage{cite}
\usepackage{multirow}
\usepackage{empheq}
\usepackage{fancyhdr}
\usepackage[caption=false, font=footnotesize]{subfig}
\SetKwInput{Input}{input}
\SetKwInput{Output}{output}

\IEEEoverridecommandlockouts
\begin{document}

	\title{Energy Efficiency of Rate-Splitting Multiple Access, and Performance Benefits over SDMA and NOMA
%		\\[-0.7ex]
		}

		\author{
			 
			\IEEEauthorblockN{Yijie Mao$^*$,  Bruno Clerckx$^\dagger$ and Victor O.K. Li$^*$ }
		\IEEEauthorblockA{$^*$The University of Hong Kong, Hong Kong, China\\
			$^\dagger$Imperial College London, United Kingdom\\
				Email: $^*$\{maoyijie, vli\}@eee.hku.hk, $^\dagger$b.clerckx@imperial.ac.uk}
			
			 \\[-3.5ex]
			\thanks{This work is partially supported by the U.K. Engineering and Physical
				Sciences Research Council (EPSRC) under grant EP/N015312/1.}	
			}

\maketitle
%\vspace{-300mm}
%\vspace{-100mm}

\thispagestyle{empty}
\pagestyle{empty}

\begin{abstract}
Rate-Splitting Multiple Access (RSMA) is a general and powerful multiple access framework for downlink multi-antenna systems, and contains Space-Division Multiple Access (SDMA) and Non-Orthogonal Multiple Access (NOMA) as special cases. RSMA relies on linearly precoded rate-splitting with Successive Interference Cancellation (SIC) to decode part of the interference and treat the remaining part of the interference as noise. Recently, RSMA has been shown to outperform both SDMA and NOMA rate-wise in a wide range of network loads (underloaded and overloaded regimes) and user deployments (with a diversity of channel directions, channel strengths and qualities of channel state information at the transmitter). Moreover, RSMA was shown to provide spectral efficiency and QoS enhancements over NOMA at a lower computational complexity for the transmit scheduler and the receivers. In this paper, we build upon those results and investigate the energy efficiency of RSMA compared to SDMA and NOMA. Considering a multiple-input single-output broadcast channel, we show that  RSMA is more energy-efficient than SDMA and NOMA in a wide range of user deployments (with a diversity
of channel directions and channel strengths). We conclude that RSMA is more spectrally and energy-efficient than SDMA and NOMA.
\end{abstract}
\maketitle

%\let\thefootnote\relax\footnotetext[1]{This work is partially supported by the U.K. Engineering and Physical
%	Sciences Research Council (EPSRC) under grant EP/N015312/1.}
%\maketitle
%\vspace{-7mm}

%\section{Introduction}
%\textit{Notations}: The boldface uppercase and lowercase letters are used to represent matrices and vectors, respectively. $\mathbb{C}$ denotes the complex space. $\mathbf{A}^H$ represents the conjugate transpose of matrix $\mathbf{A}$. $\mathbf{I}_{N}$ stands for the $N \times N$ identity matrix. 
%The closure of a set $\mathcal{S}$ is denoted by $\mathrm{cl} \;\mathcal{S}$ and its interior is denoted by $\mathring{\mathcal{S}} $.
\vspace{-0mm}
\section{Introduction}
\vspace{-0mm}
\par The cellular network is envisioned to be ultra-dense with the proliferation of the Mobile Internet and the Internet of Things (IoT). 
The corresponding energy cost is increasing rapidly and becomes a major threat for sustainable development. Much efforts have been spent to solve the problem of Energy Efficiency (EE) maximization so as to keep the optimal trade-off between the Weighted Sum Rate (WSR) and the total power consumption \cite{EEsurvey2016}. 
%The typical EE metric is defined as the ratio of the sum data rate and total power consumption that includes all the power consumed in signal processing and transmission of signal \cite{tervo2015optimalEE}. 
In \cite{tervo2015optimalEE}, the EE maximization problem in Multiple-Input Single-Output Broadcast Channel (MISO BC) subject to sum power and individual Signal-to-Interference plus Noise Ratio (SINR) constraints is investigated. The optimal EE beamformer is obtained by using the Successive Convex Approximation (SCA)-based approach and it is further extended to multi-cell in \cite{tervo2017scaRate}.  A detailed comparison of the related work on the EE problem in multi-antenna systems is illustrated in  \cite{yang2018energy}. The authors solve the EE maximization problem in Multiple-Input Multiple-Output (MIMO) BC by using the successive pseudoconvex approximation approach. All of the above works consider Space Division Multiple Access (SDMA)  based on Multi-User Linear Precoding (MU--LP) beamforming. Each receiver only decodes its intended message by fully treating any residual interference from other users as noise. However, SDMA based on MU--LP is only suited to the underloaded regime and the scenarios where the user channels are sufficiently orthogonal.  

\par 
%The past decades have witnessed the evolution of multiple access from Orthogonal Multiple Access (OMA) to Non-orthogonal Multiple Access (NOMA). 

Power-domain Non-orthogonal Multiple Access (NOMA) (simply referred to as NOMA in the sequel) based on Superposition Coding (SC) at the transmitter and Successive Interference Cancellation (SIC) at the receivers has been recognized as a promising multiple access scheme for future mobile networks \cite{noma2013saito}. Different from SDMA, some users are forced to fully decode and cancel interference from other users. Though NOMA has the capability of serving users in an overloaded regime, it is only suited to the deployments where the user channels are sufficiently aligned and exhibit a large disparity in channel strengths. Efforts have been paid to the precoder design so as to achieve the best EE in the NOMA-based MIMO BC. In \cite{EENOMAMIMO2015}, the EE maximization problem in the two-user NOMA-based MIMO BC is investigated under the assumption of only statistical Channel State Information (CSI) to be known at the transmitter.  The $K$-user NOMA-based MIMO BC is investigated in \cite{nomamimo2017}. However, the precoder is designed based on interference alignment and only power is optimized.

\par To overcome the shortcomings of SDMA and NOMA, a novel multiple access scheme called Rate-Splitting Multiple Access (RSMA) is proposed in \cite{mao2017rate}.  RSMA, based on  linear precoded Rate-Splitting (RS), has the ability to partially decode interference and partially treat interference as noise. As a consequence, RSMA bridges and unifies the two extremes of SDMA and NOMA. To partially decode interference, various messages of users are split into common and private parts  in RS. The common parts are jointly encoded and decoded by multiple users while the private parts are decoded by the corresponding users only. RSMA has been shown in \cite{mao2017rate} to be more spectrally efficient than SDMA and NOMA in a wide range of user deployments (with a diversity of channel directions and channel strengths), and in the presence of perfect and imperfect CSI at the Transmitter (CSIT). However, the EE performance of RSMA has never been studied. In the literature, the two-user RS-assisted EE maximization problem in MIMO Interference Channel (IC) has been investigated in \cite{RSEE2017AZappone} by allowing one of the two users to use RS. Different from \cite{RSEE2017AZappone}, we initiate the investigation of RS-assisted EE maximization problem in MISO BC.

\par Building upon the results in \cite{mao2017rate}, we study the EE of RSMA in this work and compare with SDMA and NOMA. To investigate the EE region achieved by different multiple access schemes, we consider a EE metric defined as the WSR divided by the total power consumption. A SCA-based beamforming algorithm is proposed to solve the RSMA EE maximization problem. SDMA and NOMA EE maximization problems are obtained as a special case of the RSMA EE maximization framework. The performance of the proposed SCA-based algorithm and the resulting EE regions achieved by different multiple access schemes are compared in the numerical results. The convergence rate of the proposed algorithm is shown to be high and the EE region of RSMA is shown to be equal to or larger than that of SDMA and NOMA in any user deployments. RSMA is therefore more energy-efficient than SDMA and NOMA.

\par The rest of the paper is organized as follows. Section \ref{sec: system model} specifies the system model and the formulated EE problem of SDMA and NOMA. The proposed EE problem of RSMA is discussed in Section \ref{sec: RS} followed by the proposed algorithm based on SCA in Section \ref{sec: algorithm}. Section \ref{sec: simulation} shows numerical results and Section \ref{sec: conclusion} concludes the paper. 

\par \textit{Notations}: $\mathbb{C}$ refers to the complex space and  $\mathbb{E}\{\cdot\}$ refers to the statistical expectation. $\mathrm{tr}(\cdot)$ is the trace.  The boldface uppercase and lowercase letters represent matrices and vectors, respectively. $\left\Vert\cdot\right\Vert$ is the Euclidean norm. 
The superscripts $(\cdot)^T$ and $(\cdot )^H$  are transpose
and conjugate-transpose operators.

%\vspace{-0mm}
\section{System model and Existing Multiple Access}
%\vspace{-0mm}
\label{sec: system model}
\subsection{System Model}
%\vspace{-0.5mm}
\par Consider the downlink transmission of a two-user\footnote{We consider a two-user scenario for readability and page constraint reasons, though the system model can be extended to the general $K$-user. The extension will be treated in the journal version of this paper.} MISO system where one Base Station (BS) equipped with $N_t$ transmit antennas serves two single-antenna users. The BS wants to transmit the messages $W_1,  W_2$ respectively to user-$1$ and user-$2$  in each time frame. The messages are encoded based on different multiple access schemes and form the transmit signal 
$\mathbf{x}\in\mathbb{C}^{N_t\times1}$. The total transmit power of the BS is subject to a power constraint $P_{t}$ as $\mathbb{E}\{\left\Vert  \mathbf{x}\right\Vert^2\}\leq P_{t}$.

\par The received signal at user-$k,\forall k\in\{1,2\}$ is
%\vspace{-1mm}
\begin{equation}
\begin{aligned}
y_{k}&=\mathbf{{h}}_{k}^{H}\mathbf{{x}}+n_{k},
\end{aligned}
\vspace{-1mm}
\end{equation}
where  $\mathbf{{h}}_{k}\in\mathbb{C}^{N_{t}\times1}$ is the channel from user-$k$ to BS. It is perfectly known at BS.
$n_{k}\sim\mathcal{CN}(0,\sigma_{n,k}^{2})$ is the Additive White Gaussian Noise (AWGN) at user-$k$ with  zero mean and variance $\sigma_{n,k}^{2}$. 

%Without loss of generality, we assume $\sigma_{n,1}^{2}=\sigma_{n,2}^{2}=1$.  Hence, the transmit Signal to Noise Ratio (SNR) is equal to the total power consumption $P_t$.
%\vspace{-0mm}
\subsection{Power Consumption Model}
%\vspace{-0.5mm}
\par The power consumption at BS contains not only the transmit power, but also the circuit power due to the electronic operations.
The linear power model specified in \cite{xu2011improving} is adopted in this work. The total power consumption is
\vspace{-0mm}
\begin{equation}
 P_{\textrm{total}}=\frac{1}{\eta}P_{\textrm{tran}}+P_{\textrm{cir}},
 \vspace{-0mm}
\end{equation}
where $P_{\textrm{tran}}\triangleq\mathbb{E}\{\left\Vert  \mathbf{x}\right\Vert^2\}$ is the transmit power. $\eta\in[0,1]$ is the  power amplifier
efficiency.  $P_{\textrm{cir}}=N_tP_{\textrm{dyn}}+P_{\textrm{sta}}$ is the circuit power. $P_{\textrm{dyn}}$ denotes the dynamic power consumption. It is  the power consumption of one active  Radio Frequency chain. $P_{\textrm{sta}}$ denotes the static power consumption, which is the power consumption of cooling systems, power supply and so on. The power consumption at user sides is omitted since the  power consumption of users is negligible
compared with the power consumption of BS  \cite{xu2011improving}.

%\vspace{-0mm}
\subsection{Existing Multiple Access}
%\vspace{-0.5mm}
\par We briefly review two baseline multiple access schemes, namely, SDMA and NOMA and the corresponding EE maximization problem. At the BS, the messages $W_1$ and  $W_2$ are independently encoded into the  data streams $s_1$ and $s_2$. The data streams
are respectively multiplied by the beamforming vectors   $\mathbf{p}_1,\mathbf{p}_2\in\mathbb{C}^{N_t\times1}$ and superposed as
\vspace{-0mm}
\begin{equation}
	\mathbf{x}=\mathbf{P}\mathbf{{s}}=\mathbf{p}_1s_1+\mathbf{p}_2s_2,
	\vspace{-0mm}
\end{equation}
where $\mathbf{{s}}\triangleq[s_1,s_2]^{T}$  and $\mathbf{P}\triangleq[\mathbf{p}_1,\mathbf{p}_2]$.
Assuming that $\mathbb{E}\{\mathbf{{s}}\mathbf{{s}}^H\}=\mathbf{I}$, the transmit power becomes $P_{\textrm{tran}}=\mathrm{tr}(\mathbf{P}\mathbf{P}^{H})$. It is
constrained by  $\mathrm{tr}(\mathbf{P}\mathbf{P}^{H})\leq P_{t}$.

\subsubsection{SDMA}
\par In the well-known MU--LP based SDMA scheme, each user only decodes its desired message by treating any residual interference as noise. The SINR at user-$k$, $\forall k\in\{1,2\}$ is given by
$
\gamma_{k}(\mathbf{P})={|\mathbf{{h}}_{k}^{H}\mathbf{{p}}_{k}|^{2}}/{(|\mathbf{{h}}_{k}^{H}\mathbf{{p}}_{j}|^{2}+N_{0,k})},
$
where $j \neq k$, $j\in\{1,2\}$. 
$N_{0,k}=W\sigma_{n,k}^{2}$ is the noise power at user-$k$ over the transmission bandwidth $W$. 
The corresponding achievable rate of user-$k$ is $R_k(\mathbf{P})= W\log_2(1+\gamma_k(\mathbf{P}))$.

\par For a given weight vector $\mathbf{u}=[u_1, u_2]$, the SDMA-based EE maximization problem is  given by
\begin{subequations}
	\begin{empheq}[left=\textrm{EE}_{\textrm{SDMA}}\empheqlbrace]{align}
		\max_{\mathbf{{P}}}\,\,&\frac{\sum_{k\in\{1,2\}}u_kR_{k}(\mathbf{P})}{\frac{1}{\eta}\mathrm{tr}(\mathbf{P}\mathbf{P}^{H})+P_{\textrm{cir}}}  \\
		\mbox{s.t.}\quad
%		&  \,\,R_{k}(\mathbf{P})\geq R_{k}^{th}, \forall k \in \{1,2\}\\
		&	\,\,\text{tr}(\mathbf{P}\mathbf{P}^{H})\leq P_{t}
	\end{empheq}
\end{subequations}
%where $R_{k}^{th}$ is the rate lowerbound of user-$k$. It ensures the Quality of Service (QoS) of each user.

\subsubsection{NOMA}
\label{sec: noma}
\par Contrary to SDMA where each user only decodes
its desired message, linearly precoded superposition coding is used in NOMA and one of the two users is required to fully decode the interfering message before decoding the desired message. SIC is deployed at user sides. The decoding order is required to be optimized together with the precoder.  $\pi$ denotes one of the decoding orders. The message of user-$\pi(1)$ is decoded before user-$\pi(2)$.  At user-$\pi(1)$, the desired message is decoded directly by treating any interference as noise. The SINR at user-$\pi(1)$ is given by
$
\gamma_{\pi(1)}(\mathbf{P})={|\mathbf{{h}}_{\pi(1)}^{H}\mathbf{{p}}_{\pi(1)}|^{2}}/{(|\mathbf{{h}}_{\pi(1)}^{H}\mathbf{{p}}_{\pi(2)}|^{2}+N_{0,\pi(1)})}.
$
At user-$\pi(2)$, the interference from user-$\pi(1)$ is decoded before decoding the desired message. The SINR at user-$\pi(2)$ to decode the message of user-$\pi(1)$  is given by 
$
\gamma_{\pi(2)\rightarrow\pi(1)}(\mathbf{P})={|\mathbf{{h}}_{\pi(2)}^{H}\mathbf{{p}}_{\pi(1)}|^{2}}/{(|\mathbf{{h}}_{\pi(2)}^{H}\mathbf{{p}}_{\pi(2)}|^{2}+N_{0,\pi(2)})}.
$
Once the message of user-$\pi(1)$ is decoded and removed from the retrieved signal via SIC, user-$\pi(2)$ decodes its intended message. The SINR experienced at user-$\pi(2)$ to decode the desired message is
$
\gamma_{\pi(2)}(\mathbf{P})={|\mathbf{{h}}_{\pi(2)}^{H}\mathbf{{p}}_{\pi(2)}|^{2}}/{N_{0,\pi(2)}}.
$  The corresponding achievable rates of user-$\pi(1)$ and user-$\pi(2)$ are $R_{\pi(1)}(\mathbf{P})= \min\{W\log_2(1+\gamma_{\pi(1)}(\mathbf{P})), W\log_2(1+\gamma_{\pi(2)\rightarrow\pi(1)}(\mathbf{P}))\} $ and $R_{\pi(2)}(\mathbf{P})=W\log_2(1+\gamma_{\pi(2)}(\mathbf{P}))$.

\par For a given weight vector $\mathbf{u}$,  the NOMA-based EE maximization problem is  given by
\begin{subequations}
	\begin{empheq}[left=\textrm{EE}_{\textrm{NOMA}}\empheqlbrace]{align}
	\max_{\pi,\mathbf{{P}}}\,\,&\frac{\sum_{k\in\{1,2\}}u_{\pi(k)}R_{\pi(k)}(\mathbf{P})}{\frac{1}{\eta}\mathrm{tr}(\mathbf{P}\mathbf{P}^{H})+P_{\textrm{cir}}}  \\
	\mbox{s.t.}\quad
%	&  \,\,R_{k}(\mathbf{P})\geq R_{k}^{th}, \forall k \in \{1,2\}\\
	&	\,\,\text{tr}(\mathbf{P}\mathbf{P}^{H})\leq P_{t}
	\end{empheq}
\end{subequations}
To maximize EE, the decoding order is required to be jointly optimized with the beamforming vectors.

\section{Problem formulation of RSMA}
\label{sec: RS}
 \par In this section, we introduce  RSMA and the formulated RSMA-based EE maximization problem. 
 When there are two users in the system, the generalized RSMA proposed in \cite{mao2017rate} reduces to the 1-layer RS investigated in \cite{RS2016hamdi}.  It is easy to extend the following formulated problem to the $K$-user 1-layer RS, 2-layer Hierarchical RS (HRS) as well as the generalized RS of \cite{mao2017rate}.
 
 \par  RSMA differs from  the existing multiple access schemes mainly due to the generation of the transmit signal $\mathbf{x}$.  The messages of both SDMA and NOMA are encoded into independent streams directly. In contrast, the message $W_k,\forall k\in\{1,2\}$ of RSMA is split into a common part $W_{k,c}$ and a private part $W_{k,p}$. The common parts of both users $W_{1,c}, W_{2,c}$ are jointly encoded into a common stream $s_c$ using a codebook shared by both users. $s_c$ is intended for both users.  The private parts are encoded into $s_1$ and $s_2$ for user-$1$ and user-$2$, respectively. The stream vector $\mathbf{s}=[s_{c},s_1,s_2]^T$ is
 linearly precoded using the beamformer $\mathbf{P}=[\mathbf{p}_{c},\mathbf{p}_{1},\mathbf{p}_{2}]$. The resulting transmit signal is
 		\vspace{-0mm}
\begin{equation}
\mathbf{x}=\mathbf{P}\mathbf{{s}}=\mathbf{p}_cs_c+\mathbf{p}_1s_1+\mathbf{p}_2s_2,
		\vspace{-0mm}
\end{equation}
The transmit power $\mathrm{tr}(\mathbf{P}\mathbf{P}^{H})$ is constrained by  $ P_{t}$ as well.
%Assuming that $\mathbb{E}\{\mathbf{{s}}\mathbf{{s}}^H\}=\mathbf{I}$, the transmit power becomes $P_{\textrm{tran}}=\mathrm{tr}(\mathbf{P}\mathbf{P}^{H})$. It is
%constrained by  $\mathrm{tr}(\mathbf{P}\mathbf{P}^{H})\leq P_{t}$.

\par The common stream $s_c$ is decoded first at both users by treating the interference from the private streams $s_1$ and $s_2$ as noise. As $s_c$ contains  part of the intended message as well as part of the message of  the interfering  user, it enables the capability of partially decoding interference and partially treating interference as noise. The SINR of decoding the common stream $s_c$  at user-$k,\forall k\in\{1,2\}$ is
\vspace{-0mm}
\begin{equation}
\gamma_{c,k}(\mathbf{P})=\frac{\left|\mathbf{{h}}_{k}^{H}\mathbf{{p}}_{c}\right|^{2}}{\left|\mathbf{{h}}_{k}^{H}\mathbf{{p}}_{1}\right|^{2}+\left|\mathbf{{h}}_{k}^{H}\mathbf{{p}}_{2}\right|^{2}+N_{0,k}}.
\vspace{-0mm}
\end{equation}
The achievable rate of decoding $s_c$ at user-$k$ is $R_{c,k}(\mathbf{P})=W\log_2(1+\gamma_{c,k}(\mathbf{P}))$. To guarantee that $s_c$ is  decoded by both users, the common rate shall not exceed
\begin{equation}
R_c(\mathbf{P})=\min\{R_{c,1}(\mathbf{P}),R_{c,2}(\mathbf{P})\}.
\end{equation}  
Note that  $R_c(\mathbf{P})$ is shared by both users. Denote $C_k$ as  the $k$th user's portion of the common rate. We have
\begin{equation}
C_1+C_2=R_c(\mathbf{P}).
\end{equation}
\par Once  $s_{c}$ is decoded and removed from the received signal via SIC, user-$k$ decodes its desired private stream $s_{k}$ by treating the interference of user-$j$ ($j\neq k$) as noise. The SINR of decoding the private stream $s_{k}$ at user-$k,\forall k\in\{1,2\}$ is
\vspace{0mm}
\begin{equation}
\label{eq: private sinr}
\gamma_{k}(\mathbf{P})=\frac{\left|\mathbf{{h}}_{k}^{H}\mathbf{{p}}_{k}\right|^{2}}{\left|\mathbf{{h}}_{k}^{H}\mathbf{{p}}_{j}\right|^{2}+N_{0,k}}.
\vspace{0mm}
\end{equation}
The achievable rate of decoding $s_k$ at user-$k$ is $R_{k}(\mathbf{P})=W\log_2(1+\gamma_{k}(\mathbf{P}))$.
The achievable rate of user-$k$ is $R_{k,tot}(\mathbf{P})=C_k+R_{k}(\mathbf{P})$. Recall from \cite{mao2017rate}, SDMA and NOMA are both sub-schemes of RSMA.

 \par Based on the above model, the RSMA-based EE maximization problem for a given weight vector $\mathbf{u}$ is
 \begin{subequations}
 	\label{eq: EE RSMA}
 	\begin{empheq}[left=\textrm{EE}_{\textrm{RSMA}}\empheqlbrace]{align}
 	\max_{\mathbf{c},\mathbf{{P}}}\,\,&\frac{\sum_{k\in\{1,2\}}u_k\left(C_k+R_{k}(\mathbf{P})\right)}{\frac{1}{\eta}\mathrm{tr}(\mathbf{P}\mathbf{P}^{H})+P_{\textrm{cir}}}  \\
 	\mbox{s.t.}\quad
% 	&  \,\,C_k+R_{k}(\mathbf{P})\geq R_{k}^{th}, \forall k \in \{1,2\} \label{constraint 3}\\
 	&  \,\,C_1+C_2\leq R_{c,k}(\mathbf{P}), \forall k \in \{1,2\}  \label{constraint 4}\\
 	&	\,\,\text{tr}(\mathbf{P}\mathbf{P}^{H})\leq P_{t} \label{constraint 5}\\
 	& \,\, \mathbf{c}\geq \mathbf{0} \label{constraint 6}
 	\end{empheq}
 	 \end{subequations}
 where $\mathbf{c}=[C_1, C_2]$ is the common rate vector required to be optimized with the beamforming vectors. Constraint (\ref{constraint 4})	ensures that the common stream can be successfully decoded at both users.

%\vspace{-1mm}
\section{SCA-based Optimization Framework}
%\vspace{-1mm}
\label{sec: algorithm}
The EE maximization problems described above are non-convex fractional programs. Motivated by the SCA algorithms adopted in the literature of EE maximization \cite{tervo2015optimalEE,He2016EE, tervo2017scaRate,Ngo2018EE}, we  propose a SCA-based beamforming algorithm to solve the formulated EE maximization problems.
To achieve a high-performance approximation, auxiliary variables are introduced to first transform the original EE problem into its equivalent problem and approximations are applied to the transformed problem iteratively.  The procedure to solve the $\textrm{EE}_{\textrm{RSMA}}$ problem will be explained next and it can be easily applied to solve the $\textrm{EE}_{\textrm{MU--LP}}$ and $\textrm{EE}_{\textrm{NOMA}}$ problems. 

First of all, by introducing scalar variables $\omega^2$,  $z$ and $t$, respectively representing the weighted sum rate, total power consumption and EE metric, problem (\ref{eq: EE RSMA}) is equivalently transformed into
\vspace{-0mm}
 \begin{subequations}
	\label{eq: EE RSMA transform}
	\begin{align}
	\max_{\mathbf{c},\mathbf{{P}},\omega, z,t}&\quad\,\, t\\	
	\mbox{s.t.}\quad	
	&\quad \frac{\omega^{2}}{z}\geq t \label{constraint 0}\\ &\sum_{k\in\{1,2\}}u_k\left(C_k+R_{k}(\mathbf{P})\right)\geq \omega^{2} \label{constraint 1}\\	
	&\quad z\geq\frac{1}{\eta}\mathrm{tr}(\mathbf{P}\mathbf{P}^{H})+P_{\textrm{cir}}  \label{constraint 2}\\	
%	&\textrm{(\ref{constraint 3}), (\ref{constraint 4}), (\ref{constraint 5}),  (\ref{constraint 6})}
	&\textrm{(\ref{constraint 4}), (\ref{constraint 5}),  (\ref{constraint 6})}
	\end{align}
\end{subequations}
The equivalence between (\ref{eq: EE RSMA transform}) and (\ref{eq: EE RSMA}) is established based on the fact that constraints  (\ref{constraint 0}), (\ref{constraint 1}) and (\ref{constraint 2}) must hold with equality at optimum.  

The difficulty of solving (\ref{eq: EE RSMA transform}) lies in the non-convexity of the constraints
% (\ref{constraint 3}), 
 (\ref{constraint 4}), (\ref{constraint 0}) and (\ref{constraint 1}). We further introduce variables  $\bm{\alpha}=[\alpha_1,\alpha_2]^T$ representing the set of the private rates. Constraints 
% (\ref{constraint 3}) and 
 (\ref{constraint 1})  is equivalent to
 		\vspace{-0mm}
 \begin{subequations}
	\begin{empheq}[left=
%	\textrm{(\ref{constraint 3}),}\,\,
	\textrm{(\ref{constraint 1})}\Leftrightarrow\empheqlbrace]{align}
	&\sum_{k\in\{1,2\}}u_{k}\left(C_{k}+\alpha_{k}\right)\geq \omega^{2} \label{con: rate non convex}\\
%	&C_{k}+\alpha_{k}\geq R_{k}^{th}, \forall k \in \{1,2\} \label{con: rate convex}\\
	&R_{k}(\mathbf{P})\geq\alpha_{k}, \forall k \in \{1,2\} \label{con: rate transform}	
   \end{empheq}
\end{subequations}

To deal with the non-convex constraint (\ref{con: rate transform}), we add variables $\bm{\vartheta}=[\vartheta_1, \vartheta_2]^T$ representing $1$ plus the SINR of each private stream. The rate constraint becomes
		\vspace{-0mm}
 \begin{subequations}
	\begin{empheq}[left=\textrm{(\ref{con: rate transform})}\Leftrightarrow\empheqlbrace]{align}
	&\vartheta_k\geq 2^{\frac{\alpha_{k}}{W}}, \forall k \in \{1,2\} \label{con: v and a}\\
	& 1+\gamma_{k}(\mathbf{P})\geq \vartheta_k , \forall k \in \{1,2\}\label{con: SINR transform}
	\end{empheq}
\end{subequations}
where (\ref{con: v and a}) is transformed from $W\log_2\vartheta_k\geq\alpha_{k}$. $\gamma_{k}(\mathbf{P})$ is calculated based on (\ref{eq: private sinr}). (\ref{con: SINR transform})  is further transformed into 
		\vspace{-0mm}
 \begin{subequations}
 	\label{con: 1+SINR}
	\begin{empheq}[left=\textrm{(\ref{con: SINR transform})}\Leftrightarrow\empheqlbrace]{align}
	&\frac{\left|\mathbf{{h}}_{k}^{H}\mathbf{{p}}_{k}\right|^{2}}{\beta_k}\geq \vartheta_k-1 , \forall k \in \{1,2\} \label{con: private SINR non-linear}\\
	&\beta_k\geq N_{0,k}+\sum_{j\neq k}\left|\mathbf{{h}}_{k}^{H}\mathbf{{p}}_{j}\right|^{2} , \forall k \in \{1,2\} \label{con: private noise interference}
	\end{empheq}
\end{subequations}
where $\bm{\beta}=[\beta_1,\beta_2]^T$ are new variables representing the interference plus noise at each user to decode its private steam. Therefore,  constraint
% (\ref{constraint 3}) and 
 (\ref{constraint 1})  can be replaced by
 		\vspace{-0mm}
\[
%\textrm{(\ref{constraint 3})},
\textrm{(\ref{constraint 1})}\Leftrightarrow 
\textrm{(\ref{con: rate non convex})},
%\textrm{(\ref{con: rate convex})},
\textrm{(\ref{con: v and a})},\textrm{(\ref{con: 1+SINR})}		\vspace{-0mm}\]

Similarly, we introduce variables $\bm{\alpha}_c=[\alpha_{c,1},\alpha_{c,2}]^H$ representing the common rate at user sides,  $\bm{\vartheta}_c=[\vartheta_{c,1}, \vartheta_{c,2}]^T$ representing $1$ plus the SINR of the common stream as well as $\bm{\beta}_{c}=[\beta_{c,1},\beta_{c,2}]^T$ representing the interference plus noise at each user to decode the common steam, constraint (\ref{constraint 4}) is equivalent to
		\vspace{-0mm}
 \begin{subequations}
	\label{con: common}
	\begin{empheq}[left=\textrm{(\ref{constraint 4})}\Leftrightarrow\empheqlbrace]{align}
	&  C_1+C_2\leq \alpha_{c,k}, \forall k \in \{1,2\}\label{con: common rate}\\
	&\vartheta_{c,k}\geq 2^{\frac{\alpha_{c,k}}{W}} , \forall k \in \{1,2\} \label{con: common SINR}\\
	&\frac{\left|\mathbf{{h}}_{k}^{H}\mathbf{{p}}_{c}\right|^{2}}{\beta_{c,k}}\geq \vartheta_{c,k}-1 , \forall k \in \{1,2\} \label{con: common SINR non-linear}\\
	&\beta_{c,k}\geq N_{0,k}+\left|\mathbf{{h}}_{k}^{H}\mathbf{{p}}_{1}\right|^{2}+\left|\mathbf{{h}}_{k}^{H}\mathbf{{p}}_{2}\right|^{2}, \forall k  \label{con: common noise interference}
	\end{empheq}
\end{subequations}
		\vspace{-0mm}

Therefore, problem (\ref{eq: EE RSMA}) is equivalently transformed into
 \[		\vspace{-0mm}
	\begin{aligned}
	\max_{\substack{\mathbf{c},\mathbf{{P}}, \omega, z,t,\\ \bm{\alpha}_{c},\bm{\alpha},\bm{\vartheta}_{c},\bm{\vartheta},\bm{\beta}_{c},\bm{\beta}}}&\quad\,\, t\\	
	\mbox{s.t.}\quad	
	&\quad \textrm{(\ref{constraint 5})}, \textrm{(\ref{constraint 6})},\textrm{(\ref{constraint 0})},\textrm{(\ref{constraint 2})}   \\
	&\quad \textrm{(\ref{con: rate non convex})},
%	\textrm{(\ref{con: rate convex})},
	\textrm{(\ref{con: v and a})},\textrm{(\ref{con: 1+SINR})},\textrm{(\ref{con: common})} 
	\end{aligned}
\]
The constraints of the transformed problem are convex except (\ref{constraint 0}), (\ref{con: private SINR non-linear}) and (\ref{con: common SINR non-linear}). 
So we use the linear approximation to approximate the non-convex part of the constraints in each iteration. 
The left side of  (\ref{constraint 0}) is approximated using the first-order lower approximation, which is given by
\vspace{-0mm}
\begin{equation}
\label{eq: approximation 1}
	\frac{\omega^{2}}{z}\geq\frac{2\omega^{[n]}}{z^{[n]}}\omega-(\frac{\omega^{[n]}}{z^{[n]}})^{2}z\triangleq\Omega^{[n]}(\omega,z),
	\vspace{-0mm}
\end{equation}
where ($\omega^{[n]},z^{[n]}$) are the values of the variables ($\omega, z$) at the output of the $n$th iteration.
The left side of  (\ref{con: private SINR non-linear}) and (\ref{con: common SINR non-linear})   are written using the linear lower bound approximation at the point ($\mathbf{p}_k^{[n]},\beta_{k}^{[n]}$) and  ($\mathbf{p}_c^{[n]},\beta_{c,k}^{[n]}$) respectively as
\vspace{-0mm}
\begin{equation}
\label{eq: approximation 2}
\begin{aligned}
\frac{\left|\mathbf{{h}}_{k}^{H}\mathbf{{p}}_{k}\right|^{2}}{\beta_{k}}&\geq{2\mathrm{Re}\left((\mathbf{{p}}_{k}^{[n]})^{H}\mathbf{{h}}_{k}\mathbf{{h}}_{k}^{H}\mathbf{{p}}_{k}\right)}/{\beta_{k}^{[n]}}\\&-\left({\left|\mathbf{{h}}_{k}^{H}\mathbf{{p}}_{k}^{[n]}\right|}/{\beta_{k}^{[n]}}\right)^{2}\beta_{k}\triangleq\Psi_{k}^{[n]}(\mathbf{{p}}_{k},\beta_{k}),
\end{aligned}
\vspace{-0mm}
\end{equation}
\vspace{-0mm} 
\begin{equation}
\label{eq: approximation 3}
\begin{aligned}
\frac{\left|\mathbf{{h}}_{k}^{H}\mathbf{{p}}_{c}\right|^{2}}{\beta_{c,k}}&\geq{2\mathrm{Re}\left((\mathbf{{p}}_{c}^{[n]})^{H}\mathbf{{h}}_{k}\mathbf{{h}}_{k}^{H}\mathbf{{p}}_{c}\right)}/{\beta_{c,k}^{[n]}}\\&-\left({\left|\mathbf{{h}}_{k}^{H}\mathbf{{p}}_{c}^{[n]}\right|}/{\beta_{c,k}^{[n]}}\right)^{2}\beta_{c,k}\triangleq\Psi_{c,k}^{[n]}(\mathbf{{p}}_{c},\beta_{c,k}).
\end{aligned}
\vspace{-0mm}
\end{equation} 

\par Based on the approximations (\ref{eq: approximation 1})--(\ref{eq: approximation 3}), problem (\ref{eq: EE RSMA}) is approximated at iteration $n$ as
\vspace{-0mm}
\begin{equation}
\label{eq: final problem}
	\begin{aligned}
\max_{\substack{\mathbf{c},\mathbf{{P}}, \omega, z,t,\\ \bm{\alpha}_{c},\bm{\alpha},\bm{\vartheta}_{c},\bm{\vartheta},\bm{\beta}_{c},\bm{\beta}}}&\quad\,\, t\\	
\mbox{s.t.}\quad	
&\quad \Omega^{[n]}(\omega,z)\geq t \\
&\quad \Psi_{k}^{[n]}(\mathbf{{p}}_{k},\beta_{k}) \geq \vartheta_{k}-1, \forall k \in \{1,2\}\\
&\quad \Psi_{c,k}^{[n]}(\mathbf{{p}}_{c},\beta_{c,k})\geq \vartheta_{c,k}-1, \forall k \in \{1,2\}\\
&\quad \textrm{(\ref{constraint 5})}, \textrm{(\ref{constraint 6})},\textrm{(\ref{constraint 2})},\textrm{(\ref{con: rate non convex})},
%\textrm{(\ref{con: rate convex})},
\textrm{(\ref{con: v and a})}, \\
&\quad\textrm{(\ref{con: private noise interference})},\textrm{(\ref{con: common rate})},\textrm{(\ref{con: common SINR})},\textrm{(\ref{con: common noise interference})}
\end{aligned}
\end{equation}
The problem (\ref{eq: final problem})  is convex and can be
solved using CVX, a package for solving disciplined convex
programs in Matlab \cite{grant2008cvx}. The SCA-based beamforming algorithm with RS is outlined in Algorithm \ref{SCA algorithm}.  In each iteration, problem (\ref{eq: final problem}) is solved and $\omega^{[n]}, z^{[n]}$, $\mathbf{P}^{[n]}, \bm{\beta}_c^{[n]}, \bm{\beta}^{[n]}$ are updated using the corresponding optimized variables. $t^{[n]}$ is the maximized EE at the output of the $n$th iteration. $\epsilon$ is the tolerance of the algorithm.

\textit{Initialization}:
% find$\{\mathbf{w}|(\ref{constraint 5})\}$
The beamformer $\mathbf{P}^{[0]}$ is initialized by finding the feasible beamformer  satisfying the transmit power constraint $(\ref{constraint 5})$. The common rate vector $\mathbf{c}^{[0]}$ is initialized by assuming the common rate $ R_{c,k}(\mathbf{P}^{[0]})$ is uniformly allocated to user-1 and user-2. $\omega^{[0]}$, $z^{[0]}$, $\beta_k^{[0]}$ and $\beta_{c,k}^{[0]}$ are initialized by replacing the inequalities of  (\ref{constraint 1}), (\ref{constraint 2}), (\ref{con: private noise interference}) and (\ref{con: common noise interference}) with equalities, respectively.

\textit{Convergence Analysis}:  As  (\ref{constraint 0}), (\ref{con: private SINR non-linear}) and (\ref{con: common SINR non-linear}) are relaxed by the first-order lower bounds (\ref{eq: approximation 1})--(\ref{eq: approximation 3}), the solution of problem (\ref{eq: final problem}) at iteration $[n]$ is also a feasible solution at iteration $[n+1]$. Therefore, the optimized objective is non-decreasing as iteration increases, $t^{[n+1]}\geq t^{[n]}$ always holds. As the EE $t$ is bounded above by the transmit power constraint \textrm{(\ref{constraint 5})}, the proposed algorithm is guaranteed to converge. Due to the linear approximation of the constraints  (\ref{constraint 0}), (\ref{con: private SINR non-linear}) and (\ref{con: common SINR non-linear}), the global optimality of the achieved solution can not be guaranteed.
 
\begin{algorithm}[t!]	
	\textbf{Initialize}: $n\leftarrow0$, $t^{[n]},\omega^{[n]}, z^{[n]}$, $\mathbf{P}^{[n]}, \bm{\beta}_c^{[n]}, \bm{\beta}^{[n]}$\;
	\Repeat{$|t^{[n]}-t^{[n-1]}|<\epsilon$}{
		$n\leftarrow n+1$\;
		Solve problem (\ref{eq: final problem}) using $\omega^{[n-1]}$, $z^{[n-1]}$, $\mathbf{P}^{[n-1]}$, $\bm{\beta}_c^{[n-1]}$, $\bm{\beta}^{[n-1]}$ and denote the optimal objective as $t^*$ and the optimal variables as $\omega^*$, $z^*$, $\mathbf{P}^*$, $\bm{\beta}_c^*$, $\bm{\beta}^*$ \;
		Update $t^{[n]}\leftarrow t^*$, $\omega^{[n]}\leftarrow\omega^*$, $z^{[n]}\leftarrow z^*$,   $\mathbf{P}^{[n]}\leftarrow \mathbf{P}^*$, $\bm{\beta}_c^{[n]}\leftarrow\bm{\beta}_c^*$,   $\bm{\beta}^{[n]}\leftarrow\bm{\beta}^*$\;				
 	}	
	\caption{SCA-based beamforming algorithm with RS}
	\label{SCA algorithm}		
\end{algorithm}
 
%	\vspace{-0mm}
\section{Numerical Results}
%	\vspace{-0mm}
\label{sec: simulation}
In this section, we evaluate the performance of RSMA by comparing the EE region of RSMA with that of SDMA and NOMA.  The two-user EE region consists of all achievable individual EE-pairs ($\textrm{EE}_1, \textrm{EE}_2$). The individual EE is defined as the individual achievable rate divided by the sum power. For example, the individual EE of user-$k$ in RSMA is 
\vspace{-0.0mm}
\begin{equation}
	\textrm{EE}_k=\frac{C_k+R_{k}(\mathbf{P})}{\frac{1}{\eta}\mathrm{tr}(\mathbf{P}\mathbf{P}^{H})+P_{\textrm{cir}}},\forall k\in\{1,2\}.
	\vspace{-0.0mm}
\end{equation}
%$\textrm{EE}_k=\frac{u_k\left(C_k+R_{k}(\mathbf{P})\right)}{\frac{1}{\eta}\mathrm{tr}(\mathbf{P}\mathbf{P}^{H})+P_{\textrm{cir}}}$.
 The boundary of the EE region is calculated by varying the weights assigned to users. 
% As the largest achievable EE region are investigated, the rate lowerbound of each user is set to 0, i.e., $R_0^{th}=0$  bit/s/Hz. 
 Following the rate region simulation in \cite{mao2017rate}, the weight of user-1 in this work is fixed to $u_1=1$ and that of user-2 is changed as $u_2=10^{[-3,-1,-0.95,\ldots,0.95,1,3]}$.
 The BS is assumed to have four transmit antennas ($N_t=4$). Without loss of generality, unit noise variance ($\sigma_{n,k}^{2}=1$) and unit bandwidth ($W=1$  Hz) is considered. The transmit power constraint is $P_t= 40$ dBm.\footnote{As the noise power is normalized, $P_t=40$ dBm also implies the transmit SNR is $10$ dB.} The static power consumption is $P_{\textrm{sta}}=30$ dBm and the power amplifier efficiency is $\eta=0.35$. 
 We follow the channel model in \cite{mao2017rate} to investigate the effect of channel angle and channel gain disparity on EE region shape. The channels are given by  $
\mathbf{h}_1=\left[1, 1, 1, 1\right]^H$,
$
\mathbf{h}_2=\gamma\times\left[
1,e^{j\theta},e^{j2\theta},e^{j3\theta}\right]^H
$.
$\gamma$ controls the channel gain disparity while $\theta$ controls the channel angle.
To simplify the notations in the results, SC--SIC is used to represent the transmission scheme of NOMA based on SC--SIC. MU--LP and RS are used to represent the transmission scheme of SDMA  and RSMA, respectively. 

\begin{figure}[t!] 
	\vspace{-1mm}
	\centering
	\includegraphics[width=2.9in]{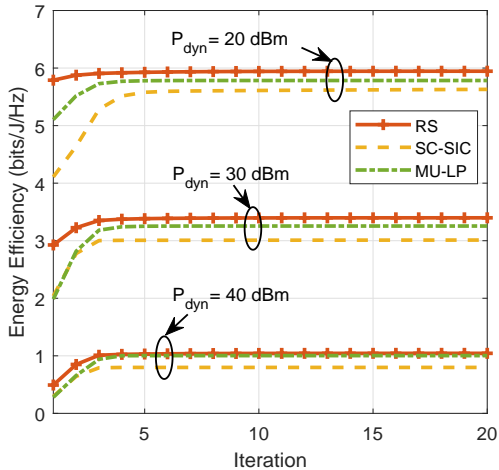}	
	\caption{ Convergence of the proposed SCA-based beamforming algorithm with different schemes, $u_1=1,u_2=1$, $\gamma=1$, $\theta=\frac{2\pi}{9}$, $P_{\textrm{dyn}}=$20, 30 and 40 dBm.}
	\label{fig: convergence}
	\vspace{-2mm}
\end{figure}

The convergence of the proposed SCA-based beamforming algorithm with RS, SC--SIC and MU--LP using one specific channel realization ($\gamma=1$, $\theta=\frac{2\pi}{9}$) are compared in Fig. \ref{fig: convergence}. The weights of the users are fixed to 1 ($u_1=u_2=1$). As mentioned in Section \ref{sec: noma}, the decoding order $\pi$ of NOMA is required to be optimized with the beamformer. For each decoding order, the SCA-based algorithm is used to solve the EE maximization problem.  Only the convergence result of the decoding order that achieves the maximal EE is illustrated in Fig. \ref{fig: convergence}.  For various dynamic power values $P_{\textrm{dyn}}$, the convergence rate of the algorithm with the three schemes are fast. All of them converge within a few iterations. However, as the SCA-based beamforming algorithm is used twice to solve the $\textrm{EE}_{\textrm{NOMA}}$ problem, the transmitter complexity of NOMA is increased comparing with MU—LP and RS. The convergence rates of all the schemes are slightly increasing as $P_{\textrm{dyn}}$ decreases. This is due to the fact that the overall optimization space is enlarged as $P_{\textrm{dyn}}$ decreases.

\begin{figure}[t!] 
	\vspace{-1mm}
	\centering
	\includegraphics[width=3.4in]{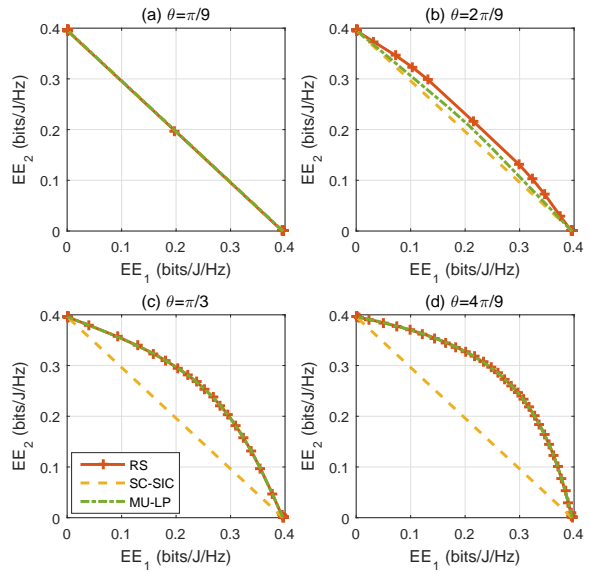}	\vspace{-3mm}
	\caption{ Achievable energy efficiency region comparison of different schemes, $\gamma=1$, $P_{\textrm{dyn}}=27$ dBm.}
	\label{fig: bias1 pcir27}
	\vspace{-3mm}
\end{figure}

\begin{figure}[t!]
    \vspace{-1mm}
	\centering
	\includegraphics[width=3.4in]{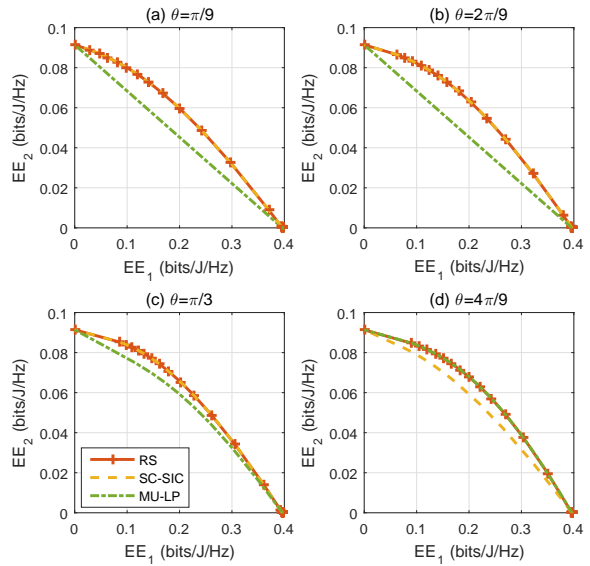}	\vspace{-3mm}
	\caption{ Achievable energy efficiency region comparison of different schemes, $\gamma=0.3$, $P_{\textrm{dyn}}=27$ dBm.}
	\label{fig: bias03 pcir27}
	\vspace{-3mm}
\end{figure}
Fig. \ref{fig: bias1 pcir27} and Fig. \ref{fig: bias03 pcir27} show the EE region comparison of different schemes when $P_{\textrm{dyn}}=27$ dBm. $\gamma$ is equal to 1 and 0.3, respectively. In all subfigures, the EE region achieved by RS is equal to or larger than that of SC--SIC and MU--LP.  In subfigure (b) of Fig. \ref{fig: bias1 pcir27}, RS outperforms SC--SIC and MU--LP. RS achieves a better EE region especially when the user channels are neither orthogonal nor aligned. As $\theta$ increases, the gap between RS and MU--LP decreases because MU--LP works well when the user channels are sufficiently orthogonal. The performance of SC--SIC becomes better when there is a 5 dB channel gain difference between users in Fig. \ref{fig: bias03 pcir27}. The EE region of SC--SIC is almost overlapped with RSMA in subfigures (a)--(c) of Fig. \ref{fig: bias03 pcir27}. However, when the user channels become sufficiently orthogonal, there is an obvious EE region improvement of RS over SC--SIC.

\begin{figure}[t!] 
		\vspace{-3mm}
	\centering
	\includegraphics[width=3.4in]{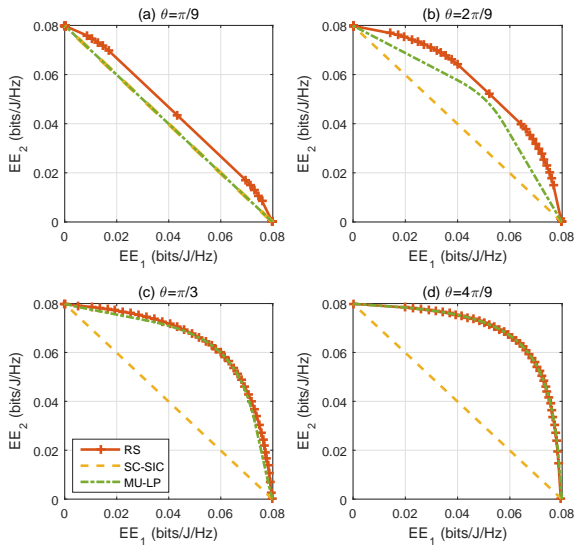}	\vspace{-4mm}
	\caption{ Achievable energy efficiency region comparison of different schemes, $\gamma=1$, $P_{\textrm{dyn}}=40$ dBm.}
	\label{fig: bias1 pcir40}
		\vspace{-1mm}
\end{figure}

\begin{figure}[t!]
		\vspace{-3mm}
	\centering
	\includegraphics[width=3.3in]{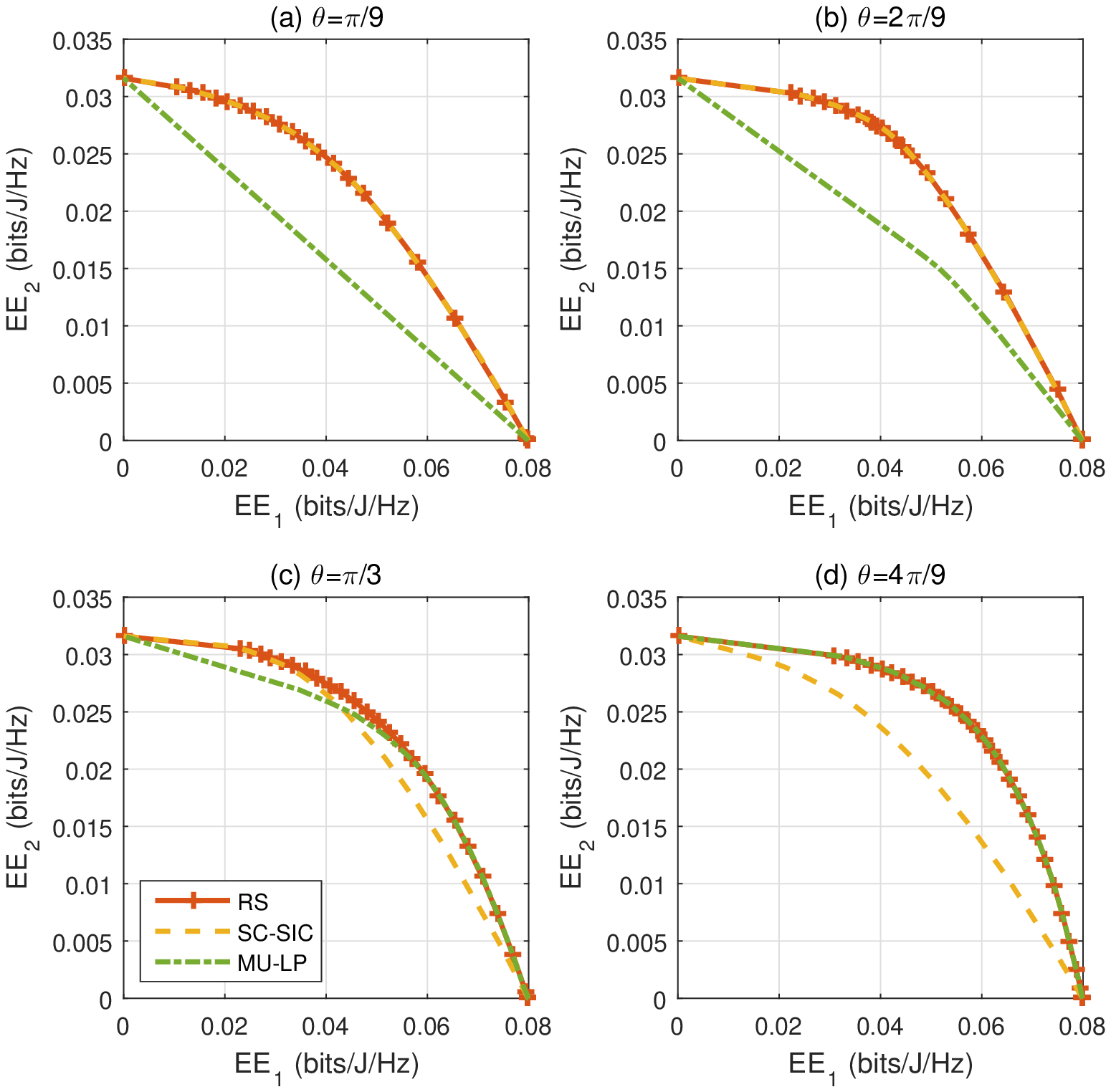}	\vspace{-2mm}
	\caption{ Achievable energy efficiency region comparison of different schemes, $\gamma=0.3$, $P_{\textrm{dyn}}=40$ dBm.}
	\label{fig: bias03 pcir40}
		\vspace{-3mm}
\end{figure}

Fig. \ref{fig: bias1 pcir40} and Fig. \ref{fig: bias03 pcir40} show the EE region comparison of different schemes  when $P_{\textrm{dyn}}=40 $ dBm. $\gamma$ is equal to 1 and 0.3, respectively. As $P_{\textrm{dyn}}$ increases, the EE regions of all multiple access schemes decrease since the denominators of individual EE increase. 
Comparing the corresponding subfigures of Fig. \ref{fig: bias1 pcir40} and Fig. \ref{fig: bias1 pcir27} (or Fig. \ref{fig: bias03 pcir40} and Fig. \ref{fig: bias03 pcir27}), RS exhibits a more prominent EE region improvement over MU--LP and SC--SIC. The EE region gaps among RS, MU--LP and SC--SIC are increasing with $P_{\textrm{dyn}}$. When $P_{\textrm{dyn}}$ is large, the circuit power dominates the total power consumption. EE is not vulnerable to the change of transmit power compared with when $P_{\textrm{dyn}}$ is small. Hence, the results of EE maximization resemble that of the WSR maximization illustrated in \cite{mao2017rate} for large $P_{\textrm{dyn}}$. In contrast,  when $P_{\textrm{dyn}}$ is small, the power of data transmission dominates the total power consumption. EE is larger when little power is used for transmission.  Interestingly, the EE regions of SC--SIC and MU--LP outperform each other at one part of the rate region while the EE region of RS is in general larger than the convex hull of SC--SIC and MU--LP regions. It clearly shows that RS softly bridges and outperforms SC--SIC and MU--LP.

%Fig. \ref{fig: bias1 pcir40 pt1} and Fig. \ref{fig: bias03 pcir40 pt1} shows the EE region comparison of different strategies  when  $P_t= 30$ dBm, $P_{\textrm{dyn}}=30 $ dBm. $\gamma$ is equal to 1 and 0.3, respectively.  Comparing with Fig. \ref{fig: bias1 pcir40} and Fig. \ref{fig: bias03 pcir40}, the rate region gap decreases as the transmit power decreases. Though the results of EE maximization is more similar to the WSR maximization when SNR decreases, the rate region improvement of RS is not obvious when SNR is small as shown in \cite{mao2017rate}. Therefore, the EE region gaps among the strategies are smaller when SNR reduces from 10 dB to 0 dB.
%\begin{figure}[h!] 
%	%	\vspace{-2mm}
%	\centering
%	\includegraphics[width=2.9in]{figure_simulation/bias1cth0Pcir40dbmPt1EEregion.eps}	\vspace{-3mm}
%	\caption{ Achievable energy efficiency region comparison of different strategies, $\gamma=1$, $P_{\textrm{dyn}}=40$ dBm, SNR=0 dB}
%	\label{fig: bias1 pcir40 pt1}
%	%	\vspace{-3mm}
%\end{figure}
%
%\begin{figure}[h!]
%	%	\vspace{-2mm}
%	\centering
%	\includegraphics[width=2.9in]{figure_simulation/bias03cth0Pcir40dbmPt1EEregion.eps}	\vspace{-3mm}
%	\caption{ Achievable energy efficiency region comparison of different strategies, $\gamma=0.3$, $P_{\textrm{dyn}}=40$ dBm, SNR=0 dB}
%	\label{fig: bias03 pcir40 pt1}
%	%	\vspace{-3mm}
%\end{figure}

\vspace{-1mm}
\section{Conclusions}
\vspace{-0mm}
\label{sec: conclusion}
To conclude, the Energy Efficiency (EE) maximization problem of RSMA in the MISO BC is investigated. An SCA-based algorithm is proposed to solve the problem. As a novel multiple access scheme, RSMA allows common symbols decoded by multiple users to be transmitted together with the private symbols decoded by the corresponding users only. It has the capability of partially decoding interference and partially treating interference as noise. Numerical results show that RSMA softly bridges and outperforms SDMA based on MU--LP and  NOMA based on SC--SIC in the realm of EE. The EE region achieved by RSMA is always equal to or larger than that achieved by SDMA and NOMA in a wide range of user deployments (with a diversity of channel directions and channel strengths). Therefore, we conclude that RSMA is not only more spectrally efficient, but also more energy efficient than SDMA and NOMA. 

\bibliographystyle{IEEEtran}
\vspace{-1mm}
\bibliography{reference}

\end{document}